\documentclass[12pt]{article}
\usepackage{a4wide,amsfonts,amsbsy,latexsym}

\def\endproof{\hbox to \hsize{\hfil $\Box$}}

\newcommand\asd{anti-self-dual}
\newcommand\sd{self-dual}

\newlength{\extraspace}
\newlength{\extraspaces}
\newcommand{\be}{\begin{equation}
\addtolength{\abovedisplayskip}{\extraspaces}
\addtolength{\belowdisplayskip}{\extraspaces}
\addtolength{\abovedisplayshortskip}{\extraspace}
\addtolength{\belowdisplayshortskip}{\extraspace}}
\newcommand{\ee}{\end{equation}}

\newcommand{\ba}{\begin{eqnarray}
\addtolength{\abovedisplayskip}{\extraspaces}
\addtolength{\belowdisplayskip}{\extraspaces}
\addtolength{\abovedisplayshortskip}{\extraspace}
\addtolength{\belowdisplayshortskip}{\extraspace}}
\newcommand{\ea}{\end{eqnarray}}

\newcommand{\baa}{\begin{array}{rcl}
\addtolength{\abovedisplayskip}{\extraspaces}
\addtolength{\belowdisplayskip}{\extraspaces}
\addtolength{\abovedisplayshortskip}{\extraspace}
\addtolength{\belowdisplayshortskip}{\extraspace}}
\newcommand{\eaa}{\end{array}}

\newcommand{\bann}{\begin{eqnarray*}
\addtolength{\abovedisplayskip}{\extraspaces}
\addtolength{\belowdisplayskip}{\extraspaces}
\addtolength{\abovedisplayshortskip}{\extraspace}
\addtolength{\belowdisplayshortskip}{\extraspace}}
\newcommand{\eann}{\end{eqnarray*}}

\newtheorem{theorem}{Theorem}[section]

\newtheorem{proposition}{Proposition}[section]
\newtheorem{lemma}{Lemma}[section]

\begin{document}

\bigskip
\centerline{{\Large\bf Hypercomplex Integrable Systems}}

\vspace{.3in}
\centerline{{\bf J.D.E. Grant, I.A.B. Strachan}}\vspace{.1in}
\centerline{Department of Mathematics, University of Hull,}
\vspace{.1in}
\centerline{Hull, HU6 7RX, England.}
\vspace{.1in}
\centerline{e-mail: j.d.grant@maths.hull.ac.uk, i.a.strachan@maths.hull.ac.uk}

\vspace{.4in}
\centerline{{\bf Abstract}}

\bigskip

\small
\hskip 10mm\parbox{5.0in}{
In this paper we study hypercomplex manifolds in four dimensions. Rather than using an
approach based on differential forms, we develop a dual approach using vector fields.
The condition on these vector fields may then be interpreted as Lax equations, exhibiting
the integrability properties of such manifolds. A number of different field equations
for such hypercomplex manifolds are derived, one of which is in Cauchy-Kovaleskaya form
which enables a formal general solution to be given. Various other properties of the
field equations and their solutions are studied, such as their symmetry properties and the
associated hierarchy of conservation laws.}
\normalsize

\bigskip

\tableofcontents

\section{Introduction}

The study of hyperK\"ahler geometries has developed in two distinct directions. Starting with
the work of Calabi \cite{C} there has been a purely geometric vein, where such manifold are
constructed and studied geometrically without reference to any defining set of
field equations. The second vein starts with such field equations - systems of differential
equations, and uses solutions of such systems to construct the manifolds.

\medskip

To show how these two approaches are connected it is necessary to restrict
one's attention to four dimensions, where the hyperK\"ahler condition is
equivalent to the existence of a metric with \asd\ Weyl tensor and vanishing Ricci
tensor. Such metrics were shown by Penrose to have a corresponding twistor space,
and conversely, that from such a twistor space one may reconstruct the metric.
Such a structure appears at first sight to be special to four dimensions where there
is the notion of self-duality, but it was shown in \cite{Sa} that hyperK\"ahler metric have
corresponding twistor space in any $4N$-dimensional space.

\medskip

With more recent work, initiated by Ward \cite{W} (see also \cite{MW}), these two threads may be
seen to be intimately interwoven. The existence of a (suitable) twistor space indicates
that one is dealing with an integrable system. Thus any set of field equations for
a hyperK\"ahler metric provides one with an example of a multidimensional integrable
system. From a solution to this integrable system one may construct the associated
twistor space (and vica-versa), whose properties may be studied, and hence properties
of the metric, without recourse to the particular differential
equation, whose precise form depends on the particular coordinate
representation being used.

\medskip

The study of hypercomplex manifolds has had a similar history. In \cite{FP}
Finley and Pleba\'nski
studied field equations which were derived from the following condition on the
self-dual 2-form $\Sigma^i\,,$
\[
d \boldsymbol{\Sigma}^i =
\boldsymbol{\alpha} \wedge \boldsymbol{\Sigma}^i\,,\qquad\qquad i = 1\,,2\,,3\,,
\]
though without mention of the associated complex geometry. Similarly Boyer \cite{Bo} studied the
geometric aspects, but did not write down field equations for such metrics. In neither
case was the link with integrable systems made.
The aim of this paper is to show that one may write
down systems of differential equations, solution of which define hypercomplex
manifolds (here we will restrict out attention to four dimensions, though
many of the ideas will generalize to $4N$-dimensions). Since hypercomplex
manifolds also have associated twistor spaces (in four dimensions the hypercomplex condition
implies, though is not implied by, the anti-self-duality of the Weyl tensor \cite{Bo})
the field equations for these systems will be examples
of multicomponent, multidimensional integrable systems.

\medskip

Since hyperK\"ahler metrics are obviously K\"ahler they may be written in terms of
a single functions, the K\"ahler potential $\Omega\,:$
\[
{\mathbf{g}} =
\frac{\partial^2 \Omega}{\partial x^i \partial{\tilde{x}}^j} \,dx^i\,d{\tilde{x}}^y\,,
\qquad i\,,j=1\,2\,.
\]
In four dimensions the hyperK\"ahler
conditions results in the differential equation

\[
\Omega_{x^1 {\tilde{x}}^1}\Omega_{x^2{\tilde{x}}^2}-
\Omega_{x^1 {\tilde{x}}^2}\Omega_{x^2{\tilde{x}}^1}=1
\]
known as Pleba\'nski's equation. In this form it is hard (though not impossible)
to apply ideas from the theory of integrable systems, which are best suited to
evolutionary type equations. In \cite{G1} the first author showed how, by performing
a suitable Legendre transformation one may obtain an equation in evolutionary,
or Cauchy-Kovaleskaya form, namely

\begin{equation}
\psi_{tt} = \psi_{zt} \psi_{xy} - \psi_{yz} \psi_{xt}
\label{grant}
\end{equation}
and the second author \cite{S1} showed how to construct the associated integrable
hierarchy, based on the study of the generalised symmetries of this equation.

\medskip

In this paper the following two-component generalization of (\ref{grant}) will
be studied:

\begin{equation}
\begin{array}{rcl}
g_{tt} & = & \{g_x, g\} + \{h, g_z\}\,, \\
h_{tt} & = & \{h_x, g\} + \{h, h_z\}\,.
\end{array}
\label{evol}
\end{equation}
From the purely integrable systems aspect, considered in section 3, one may view this system
as resulting from dropping the volume preserving condition on the
vector fields which appear in the Lax pair used to construct (\ref{grant}).  This
approach will be followed in section 3. However, the system
has a more geometric interpretation: solutions define hypercomplex
metrics. This aspect will be considered in the next section. It should be
pointed out that these hypercomplex structure are not the most general
possible, but a particular subclass. The more general case will be
considered in a future paper.

\medskip

\section{Geometrical Description}

\subsection{Hypercomplex geometry}

In this subsection, we wish to investigate a special subclass 
of four-dimensional hyper-complex structures. In particular, 
assume we are working on a four-manifold $X$, and that 
we have a local basis for the tangent space, in the form 
of a set of four linearly-independent vector fields, 
$\{ \mathbf{e}_i: i = 1, \dots , 4 \}$. We wish to study 
the geometrical structures which arise if we assume that 
these vector fields obey the Lie-bracket relations:
\begin{equation}
\begin{array}{rcl}
\left[ \mathbf{e}_1, \mathbf{e}_2 \right] +
\left[ \mathbf{e}_3, \mathbf{e}_4 \right] &=& 0,
\\
\left[ \mathbf{e}_1, \mathbf{e}_3 \right] +
\left[ \mathbf{e}_4, \mathbf{e}_2 \right] &=& 0,
\\
\left[ \mathbf{e}_1, \mathbf{e}_4 \right] +
\left[ \mathbf{e}_2, \mathbf{e}_3 \right] &=& 0.
\end{array}
\label{alg}
\end{equation}
We will show that in this case, the manifold 
$X$ is (locally) hyper-complex. Recall that 
a manifold $M$ of dimension $n=4m$ 
is hyper-complex if it admits three integrable 
complex structures, $\mathbf{I}, 
\mathbf{J}, \mathbf{K}$, which 
obey the quaternion multiplication relations:
\begin{eqnarray*}
&\mathbf{I}^2 = \mathbf{J}^2 = \mathbf{K}^2 = - \mbox{Id}_{T_x M},&
\\
&\mathbf{I} \circ \mathbf{J} = \mathbf{K}, \qquad
\mathbf{J} \circ \mathbf{K} = \mathbf{I}, \qquad
\mathbf{K} \circ \mathbf{I} = \mathbf{J}.&
\label{quat}
\end{eqnarray*}
Such structures imply that the bundle of linear frames $L(M)$ 
reduces from a $GL(4m, \mathbb{R})$ bundle to a $GL(m, \mathbb{H})$ 
bundle. In the special case of four dimensions, the frame bundle 
reduces to a $GL(1, {\mathbb{H}}) \cong {\mathbb{H}}^* \cong {\mathbb{R}}^+
\times SU(2)$ bundle. Since $SU(2)$ is a subgroup of $SO(4)$, this means
that in four dimensions, a hypercomplex structure automatically defines
a conformal structure. In particular, there is a metric, unique up to a
scale, with respect to which all the complex structures are Hermitian:
\[
\mathbf{g} (\mathbf{IX}, \mathbf{IY}) = 
\mathbf{g} (\mathbf{JX}, \mathbf{JY}) 
= \mathbf{g} (\mathbf{KX}, \mathbf{KY})
= \mathbf{g} (\mathbf{X}, \mathbf{Y}),
\label{herman}
\] 
where $\mathbf{X}, \mathbf{Y}$ are arbitrary sections of $TM$. The
structures $\mathbf{I}, \mathbf{J}, \mathbf{K}$ along with a
representative metric in the conformal structure, ${\mathbf{g}}\,,$ define a 
hyper-hermitian structure. Each metric in 
this conformal structure has anti-self-dual Weyl tensor, 
with respect to the canonical orientation defined by any of the complex 
structures.

\medskip

In the case we wish to study, we have a set of vector fields which
satisfy the relations of equation (\ref{alg}). If we define the dual 
basis $\{ \boldsymbol{\epsilon}^i \}$ for $T^* X$, then we will show that the 
vector fields define three integrable complex structures which obey the
relations given in equation (\ref{quat}) and that the metrics with
respect to which all these structures are hermitian are conformal to
the metric:
\[
\mathbf{g} = 
\boldsymbol{\epsilon}^1 \otimes \boldsymbol{\epsilon}^1
+ \boldsymbol{\epsilon}^2 \otimes \boldsymbol{\epsilon}^2
+ \boldsymbol{\epsilon}^3 \otimes \boldsymbol{\epsilon}^3
+ \boldsymbol{\epsilon}^4 \otimes \boldsymbol{\epsilon}^4.
\]

The plan of the section is as follows. In the next section, we 
prove the assertions made above concerning the existence of 
hyper-complex structures, and a compatible conformal structure.
Next the relation between this approach, based on vector fields,
and the more usual approach based on forms is given. The vector
field approach is useful for two reasons: firstly it enables
field equations to be derived easily, and secondly it makes the
connection with integrable systems more transparent.
We then consider various coordinate versions of these equations. 

\medskip

\begin{theorem}
On a four-dimensional manifold $X$, if there exist 
four linearly-independent, 
non-vanishing vector fields 
$\{\mathbf{e}_i: i = 1, 2, 3, 4\}$ 
on a manifold $X$ which 
obey the Lie Bracket relations
\begin{equation}
\begin{array}{rcl}
\left[ \mathbf{e}_1, \mathbf{e}_2 \right] +
\left[ \mathbf{e}_3, \mathbf{e}_4 \right] &=& 0,
\\
\left[ \mathbf{e}_1, \mathbf{e}_3 \right] +
\left[ \mathbf{e}_4, \mathbf{e}_2 \right] &=& 0,
\\
\left[ \mathbf{e}_1, \mathbf{e}_4 \right] +
\left[ \mathbf{e}_2, \mathbf{e}_3 \right] &=& 0,
\end{array}
\label{alg2}
\end{equation}
then the manifold is locally hypercomplex. Moreover, there exists a
unique conformal structure, defined by the representative metric:
\begin{equation}
\mathbf{g} = 
\boldsymbol{\epsilon}^1 \otimes \boldsymbol{\epsilon}^1
+ \boldsymbol{\epsilon}^2 \otimes \boldsymbol{\epsilon}^2
+ \boldsymbol{\epsilon}^3 \otimes \boldsymbol{\epsilon}^3
+ \boldsymbol{\epsilon}^4 \otimes \boldsymbol{\epsilon}^4.
\label{metric}
\end{equation}
with respect to which all of the complex structures are Hermitian. 
\end{theorem}

\bigskip

\noindent{\bf Proof} Consider maps $(\mathbf{I},
\mathbf{J}, \mathbf{K})$ on $TX$ defined by 
\[
\begin{array}{rclrclrclrcl}
\mathbf{I}  ( \mathbf{e}_1 )& = &+ \mathbf{e}_2, 
&\mathbf{I} ( \mathbf{e}_2 )& = &- \mathbf{e}_1, 
&\mathbf{I} ( \mathbf{e}_3 )& = &+ \mathbf{e}_4, 
&\mathbf{I} ( \mathbf{e}_4 )& = &- \mathbf{e}_3, 
\\
\mathbf{J}  ( \mathbf{e}_1 )& = &+ \mathbf{e}_3, 
&\mathbf{J} ( \mathbf{e}_2 )& = &- \mathbf{e}_4, 
&\mathbf{J} ( \mathbf{e}_3 )& = &- \mathbf{e}_1, 
&\mathbf{J} ( \mathbf{e}_4 )& = &+  \mathbf{e}_2, 
\\
\mathbf{K}  ( \mathbf{e}_1 )& = &+ \mathbf{e}_4, 
&\mathbf{K} ( \mathbf{e}_2 )& = &+ \mathbf{e}_3, 
&\mathbf{K} ( \mathbf{e}_3 )& = &- \mathbf{e}_2, 
&\mathbf{K} ( \mathbf{e}_4 )& = &- \mathbf{e}_1. 
\end{array}
\]
To impose that these structures are integrable, we consider the
complexification of the tangent space $T_c X = T X \otimes \mathbb{C}$. 
Consider first the structure $\mathbf{I}$. The almost complex structure 
leads to a direct sum decomposition of $T_c X$ 
as $T^{(1, 0)} \oplus T^{(0, 1)}$ where ${\mathbf{v}} \in T^{(1, 0)}$ if
$\mathbf{I} (\mathbf{v}) = i \mathbf{v}$, and ${\mathbf{v}} \in T^{(0, 1)}$ if
${\mathbf{I}} ({\mathbf{v}}) = - i {\mathbf{v}}$. The complex structure is
integrable iff the space $T^{(1, 0)}$ is closed under the Lie bracket. 
In the case of the structure $\mathbf{I}$, $T^{(1, 0)}$ is spanned by the
vector fields $\{ {\mathbf{e}}_1 + i {\mathbf{e}}_2, {\mathbf{e}}_3 + i
{\mathbf{e}}_4 \}$, and the only non-trivial Lie bracket we must consider
is
\[
\left[ \mathbf{e}_1 + i \mathbf{e}_2, \mathbf{e}_3 + i
\mathbf{e}_4 \right] = 
\left[ \mathbf{e}_1 , \mathbf{e}_3 \right] 
+ \left[ \mathbf{e}_4, \mathbf{e}_2 \right]
+ i \left( \left[ \mathbf{e}_1 , \mathbf{e}_4 \right] 
+ \left[ \mathbf{e}_2, \mathbf{e}_3 \right] \right). 
\]
For the right hand side to lie in $T^{(1, 0)}$ implies a number of
algebraic relations which, together with the analogous equations for
$\mathbf{J}$ and $\mathbf{K}$ and the quaternionic relations, implies
that
\[
\begin{array}{rcl}
\left[ \mathbf{e}_1, \mathbf{e}_2 \right] +
\left[ \mathbf{e}_3, \mathbf{e}_4 \right] &=&
-A_2 {\mathbf{e}}_1+A_1 {\mathbf{e}}_2-A_4 {\mathbf{e}}_3+A_3 {\mathbf{e}}_4,
\\
\left[ \mathbf{e}_1, \mathbf{e}_3 \right] +
\left[ \mathbf{e}_4, \mathbf{e}_2 \right] &=&
-A_3 {\mathbf{e}}_1+A_4 {\mathbf{e}}_2+A_1 {\mathbf{e}}_3-A_2 {\mathbf{e}}_4,
\\
\left[ \mathbf{e}_1, \mathbf{e}_4 \right] +
\left[ \mathbf{e}_2, \mathbf{e}_3 \right] &=&
-A_4 {\mathbf{e}}_1-A_3 {\mathbf{e}}_2+A_2 {\mathbf{e}}_3+A_1 {\mathbf{e}}_4
\end{array}
\]
for some set of functions $\{A_1,A_2,A_3,A_4\}\,.$ Thus by construction the
corresponding metric (\ref{metric}) is hypercomplex.
It will be useful in what follows to combine these functions into a one form
${\mathbf{A}}=A_i \boldsymbol{\epsilon}^i\,$ and regard this as a connection on $X\,.$

\bigskip

Unlike the conditions for a metric to be hyperK\"ahler, the hypercomplex conditions
are invariant under conformal changes of the metric. If
${\mathbf{g}}\rightarrow e^{2\Lambda} {\mathbf{g}}$ then the corresponding transformation
for the connection ${\mathbf{A}}$ is
\[
{\mathbf{A}} \rightarrow{\mathbf{A}}+2d\Lambda\,.
\]
The subclass of hypercomplex structures which we are examining in this paper
are defined by the conformal invariant condition $d{\mathbf{A}}=0\,,$ i.e. the
connection defined by ${\mathbf{A}}$ is flat.
Thus locally one may define, for this subclass of hypercomplex structures,
the conformal factor so that ${\mathbf{A}}=0\,.$ This fixes the conformal
structure.

\bigskip

It is a straightforward exercise using the explicit form of the complex
structures given above to show that metric $\mathbf{g}$ given above is
Hermitian with respect to each of the complex structures, and that it is
the unique symmetric tensor (up to a rescaling) with this property.

\endproof

\bigskip

The hypercomplex condition determines a metric up to a conformal factor. The
zero curvature condition on the connection $\mathbf{A}$ may be used to fix
the conformal factor. This determines the metric uniquely, up to trivial
transformations.
In terms of a null tetrad in which
\[
\mathbf{g} = 
\boldsymbol{\epsilon}^1 \otimes_S \boldsymbol{\epsilon}^2+
\boldsymbol{\epsilon}^3 \otimes_S \boldsymbol{\epsilon}^4
\]
the conditions (\ref{alg2}) become
\begin{equation}
\begin{array}{rcl}
\left[\mathbf{e}_1,\mathbf{e}_2\right]+\left[\mathbf{e}_3,\mathbf{e}_4\right] & = & 0 \,, \\
\left[\mathbf{e}_1,\mathbf{e}_3\right] & = & 0 \,, \\
\left[\mathbf{e}_2,\mathbf{e}_4\right] & = & 0 \,.
\end{array}
\label{alg2null}
\end{equation}
This form will be used in later sections and is also used in the following example.

\bigskip

\noindent{\bf Example} Consider the following vector fields

\[
\begin{array}{rclcrcl}
{\mathbf{e}_1} & = & \partial_w & \qquad &{\mathbf{e}_2} & = & 
(1+|w|^2) \partial_{\overline{w}} + {\overline{z}}w\partial_{\overline{z}} \,, \\
{\mathbf{e}_4} & = & \partial_z & \qquad &{\mathbf{e}_3}& = &
(1+|z|^2) \partial_{\overline{z}} + z{\overline{w}}\partial_{\overline{w}}\,.
\end{array}
\]
It is easy to verify that these satisfy the conditions (\ref{alg2null}). Hence, by the
above theorem, the corresponding Hermitian hypercomplex metric on ${\mathbb{C}}^2$ is
conformal to
\[
{\mathbf{g}} = (1+|z|^2) dw\,d{\overline{w}}+(1+|z|^2) dz\,d{\overline{z}}
- z{\overline{w}}\, d{\overline{z}} - {\overline{z}}w dz\,d{\overline{w}}\,.
\]
Note, this metric is locally conformal to the Fubini-Study metric on ${\mathbb{CP}}^2\,.$
However this construction does not extend from ${\mathbb{C}}^2$ to ${\mathbb{CP}}^2\,.$ If it
did it would contradict Boyer's \cite{Bo} classification of compact hyperHermitian manifolds.
Other examples, with tri-holomorphic Killing vectors, have been constructed in \cite{CTV,GT}.

\bigskip

For the metric to be hyperK\"ahler the corresponding
K\"ahler forms defined by
\begin{eqnarray*}
{\mathbf{\Omega}}_I (X,Y) & = & {\mathbf{g}}(IX,Y)\,,\\
{\mathbf{\Omega}}_J (X,Y) & = & {\mathbf{g}}(JX,Y)\,,\\
{\mathbf{\Omega}}_K (X,Y) & = & {\mathbf{g}}(KX,Y)\,,
\end{eqnarray*}
must be closed, or equivalently, that $\nabla I = \nabla J = \nabla K =0\,.$
It was shown in \cite{MN} that these conditions are equivalent to the
vector fields ${\mathbf{e}}_i$ being volume preserving, that is
\[
{\cal L}_{\mathbf{e_i}} {\boldsymbol{\omega}} = 0
\]
where $\boldsymbol{\omega}$ is some volume form.

\subsection{Dual Description}

The starting point for the study of four dimensional hypercomplex manifolds
has traditionally been the equation

\begin{equation}
d \boldsymbol{\Sigma}^i =
\boldsymbol{\alpha} \wedge \boldsymbol{\Sigma}^i\,,\qquad\qquad i = 1\,,2\,,3
\label{form}
\end{equation}
where the $\boldsymbol{\Sigma}^i$ are self-dual two forms on the manifold $X\,.$ In this
paper we have so far used a dual description, using vector fields rather than
forms. This subsection is intended to bridge the gap between these two
approaches.
It is first necessary to fix some notation.
The connection one-forms are defined by

\[
d\boldsymbol{\epsilon}^i + \boldsymbol{\Gamma}^i_{~j}\wedge \boldsymbol{\epsilon}^j = 0 \,,
\]
so, in components, $\boldsymbol{\Gamma}^i_{~j} = \Gamma^i_{~jk}\boldsymbol{\epsilon}^k\,.$
The antisymmetric parts of $\Gamma^i_{~jk}$ are related to the structure
functions defined by the Lie bracket
$[{\mathbf{e}}_j,{\mathbf{e}}_k]=c_{jk}^{~~i}{\mathbf{e}}_i\,$ by
\[
\Gamma^i_{~[jk]} = \frac{1}{2} c_{jk}^{~~i}\,.
\]
This formula enable one to connect these two approaches.

\bigskip

\begin{proposition}

The connection between equations (\ref{metric}) and (\ref{form}) is given by the
formulae

\[
\boldsymbol{\alpha} = {\mathbf{A}}-\boldsymbol{\chi}
\]
where
\begin{eqnarray*}
{\mathbf{A}} & = & A_i \boldsymbol{\epsilon}^i \,, \\
{\boldsymbol{\chi}} & = & c_{ij}^{~~j} \boldsymbol{\epsilon}^i\,.
\end{eqnarray*}
\end{proposition}

\bigskip

\noindent{\bf Proof} Consider the self-dual two-form
$\boldsymbol{\Sigma}^1 = 
\boldsymbol{\epsilon}^1 \wedge\boldsymbol{\epsilon}^2+
\boldsymbol{\epsilon}^3 \wedge\boldsymbol{\epsilon}^4\,.$
Then
\begin{eqnarray*}
d \boldsymbol{\Sigma}^1 & = & 
d(\boldsymbol{\epsilon}^1 \wedge\boldsymbol{\epsilon}^2+
\boldsymbol{\epsilon}^3 \wedge\boldsymbol{\epsilon}^4) \\
& = &
-\Gamma^1_{~[ab]}
\boldsymbol{\epsilon}^a \wedge\boldsymbol{\epsilon}^b \wedge \boldsymbol{\epsilon}^2 
+\Gamma^2_{~[ab]}
\boldsymbol{\epsilon}^a \wedge\boldsymbol{\epsilon}^b \wedge \boldsymbol{\epsilon}^1
-\Gamma^3_{~[ab]}
\boldsymbol{\epsilon}^a \wedge\boldsymbol{\epsilon}^b \wedge \boldsymbol{\epsilon}^4 
+\Gamma^4_{~[ab]}
\boldsymbol{\epsilon}^a \wedge\boldsymbol{\epsilon}^b \wedge \boldsymbol{\epsilon}^3
\\
& = & -\frac{1}{2} c^{~~1}_{ab} 
\boldsymbol{\epsilon}^a \wedge\boldsymbol{\epsilon}^b \wedge \boldsymbol{\epsilon}^2
+\frac{1}{2} c^{~~2}_{ab}
\boldsymbol{\epsilon}^a \wedge\boldsymbol{\epsilon}^b \wedge \boldsymbol{\epsilon}^1
-\frac{1}{2} c^{~~3}_{ab} 
\boldsymbol{\epsilon}^a \wedge\boldsymbol{\epsilon}^b \wedge \boldsymbol{\epsilon}^4
+\frac{1}{2} c^{~~4}_{ab}
\boldsymbol{\epsilon}^a \wedge\boldsymbol{\epsilon}^b \wedge \boldsymbol{\epsilon}^3\\
& = & +\boldsymbol{\epsilon}^1\wedge\boldsymbol{\epsilon}^2\wedge
(c_{ab}^{~~a} \boldsymbol{\epsilon}^b + A_3\boldsymbol{\epsilon}^3+A_4\boldsymbol{\epsilon}^4)
+\boldsymbol{\epsilon}^3\wedge\boldsymbol{\epsilon}^4\wedge
(c_{ab}^{~~a} \boldsymbol{\epsilon}^b + A_1\boldsymbol{\epsilon}^1+A_2\boldsymbol{\epsilon}^2)\\
& = & (\boldsymbol{\epsilon}^1 \wedge\boldsymbol{\epsilon}^2+
\boldsymbol{\epsilon}^3 \wedge\boldsymbol{\epsilon}^4) \wedge
( {\mathbf{A}}-\boldsymbol{\chi})\,,\\
& = & ( {\mathbf{A}}-\boldsymbol{\chi})\wedge\boldsymbol{\Sigma}^1\,.\\
\end{eqnarray*}
The manipulations for the remaining self-dual forms are identical

\endproof

\bigskip

\noindent The hypercomplex condition on the metric is invariant under conformal changes
of the metric. Thus one needs to calculate the transformation properties of these forms
under such a change.

\begin{lemma}
Under the conformal change ${\mathbf{g}}\rightarrow e^{2\Lambda}{\mathbf{g}}$ the
forms $\boldsymbol{\alpha}\,,{\mathbf{A}}\,,\boldsymbol{\chi}$ transform as

\begin{eqnarray*}
\boldsymbol{\alpha}&\rightarrow&\boldsymbol{\alpha} + 2 d \Lambda\,, \\
{\mathbf{A}}&\rightarrow&{\mathbf{A}} - d \Lambda\,,\\
\boldsymbol{\chi}&\rightarrow&\boldsymbol{\chi} - 3 d \Lambda\,.
\end{eqnarray*}

\end{lemma}

\bigskip

\noindent{\bf Proof} The proof of this is entirely straightforward, following directly
from the definition of the various forms. It is interesting to note that
from these one may construct two linearly independent conformally invariant
two forms.

\endproof

\bigskip
Locally, if $d\boldsymbol{\alpha}=0$ then a conformal factor may be found
so the resulting metric is hyperK\"ahler. The metrics studied in this paper
come from the conformally invariant condition $d{\mathbf{A}}=0\,.$ This
leaves another conformally invariant condition $d\boldsymbol{\chi}=0\,,$ the
significance of which will be investigated elsewhere.

\subsection{Local coordinate representations}

The aim of this section is to give some local coordinate descriptions
of the equations (\ref{alg2}). The main point that we make use of is
that the equations (\ref{alg2}) may be interpreted as the integrability
conditions for null planes in the complexified tangent space of the
manifold, and that we can introduce coordinates which naturally describe
these surfaces.

\medskip

With this approach in mind, we consider the complexified tangent bundle,
$T_c X = TX \otimes \mathbb{C}$. From the vector fields $\{ \mathbf{e}_i
\}$, we define the complex basis for $T_c X$:
\bann
\mathbf{u} &=& \mathbf{e}_1 + i \mathbf{e}_2,\\
\mathbf{v} &=& \mathbf{e}_1 - i \mathbf{e}_2,\\
\mathbf{w} &=& \mathbf{e}_3 + i \mathbf{e}_4,\\
\mathbf{x} &=& \mathbf{e}_3 - i \mathbf{e}_4.
\label{cvecs}
\eann
In the case where we are considering the complexification of a real
Riemannian structure, we would impose the reality condition that 
$\mathbf{u}$ and $\mathbf{v}$ would be related by complex conjugation,
as would $\mathbf{w}$ and $\mathbf{x}$. More generally, we require the
existence of a real structure on $T_c X$ to define real slices of
different signatures. Since we will be interested in metrics of both
Riemannian and ultra-hyperbolic signatures, we will generally assume,
for the moment, that the vector fields $\mathbf{u}, \mathbf{v},
\mathbf{w}, \mathbf{x}$ are complex, and simply impose the relevant
reality conditions later.

\medskip

In terms of the vector fields $\mathbf{u}, \mathbf{v},
\mathbf{w}, \mathbf{x}$, the relations (\ref{alg2}) take the form:
\bann
&\left[ \mathbf{u}, \mathbf{w} \right] = 0,&
\\
&\left[ \mathbf{u}, \mathbf{v} \right] + 
\left[ \mathbf{w}, \mathbf{x} \right] = 0,&
\\
&\left[ \mathbf{v}, \mathbf{x} \right] = 0.&
\eann
The first of these equations states that at each point (on a suitable
local neighbouhood of) $X$, the plane spanned by the vector
fields $\mathbf{u}$ and $\mathbf{w}$ is integrable. It follows
that we can introduce coordinates $(t, y)$ on each of
these surfaces with 
\[
{\mathbf{u}} = {\partial}_t, \qquad 
{\mathbf{w}} = {\partial}_y.
\]
If we now consider the space with $t$ and $y$ held constant, the last of
equations (\ref{compalg}) tells us that this space forms an integrable
plane as well. As such, we can introduce coordinates $(x, z)$ on these
planes, and the coordinates $(t, x, y, z)$ should give a suitable
coordinate system on some local region of $X$. Note that the coordinates
$(x, z)$ are not fully determined by these conditions. The different 
choices of coordinates correspond to different coordinate expressions 
for the vector fields $\mathbf{v}$ and $\mathbf{x}$, and in fact there
are several geometrically distinct coordinate systems that we wish to
investigate:

\subsubsection{Case I}

The first case we consider is where we take the coordinates $(x, z)$ to
be null. This is the analogue of the usual complex coordinate
description of Kahler metrics. Using the second of equations
(\ref{compalg}), we introduce functions $a$ and $b$ 
and let $\mathbf{v}$ and $\mathbf{x}$ take the form
\[
{\mathbf{v}} = a_y {\partial}_x - b_y {\partial}_z, 
\qquad 
{\mathbf{x}} = - a_t {\partial}_x + b_t {\partial}_z.
\]
This is the generalisation of the form used 
in \cite{CMN} for the case of half-flat metrics. 
The functions $a$ and $b$ must now satisfy the equation
\begin{equation}
\{a, a_x\} = \{b, a_z\}, \qquad \{a, b_x\} = \{b, b_z\},
\label{heaven}
\end{equation}
where we have defined the Poisson Bracket by
\[
\{f, g\} = f_t g_y - f_y g_t.
\]
Equations (\ref{heaven}) are a generalisation of the 
First Heavenly equation which describes half-flat 
metrics \cite{Pl}. To see this, we note that the 
vectors in (\ref{alg2}) define a metric which is 
conformal to a half flat metric if the vectors 
$\mathbf{e}_i$ are divergence free with respect to 
some volume element $\boldsymbol{\omega}$ \cite{MN}. 
Taking ${\boldsymbol{\omega}} = dt \wedge dx \wedge dy \wedge dz$, we find 
that there exists a function $\Omega$ such that 
$a = {\Omega}_z, b = {\Omega}_y$. We can therefore 
integrate equations (\ref{heaven}) once and rescale 
our coordinates $(y, z)$ such that $\Omega$ 
satisfies the equation
\[
\{ \Omega_y, {\Omega}_z \} = 1,
\]
which is the First Heavenly equation \cite{Pl}.

In this case, we may reconstruct the metric, $\mathbf{g}$, which takes
the form:
\[
{\mathbf{g}} = 4 \left( b_t a_y - a_t b_y \right)^{-1} 
\left[ dt \otimes \left( a_t dz + b_t dx \right) +
dy \otimes \left( a_y dz + b_y dx \right) \right].
\]
We see that in this form, the coordinates $(t, x, y, z)$ are all null,
and are tailored to the geometry of the spheres worth of integrable null
planes through each point of the space. 

The metric is Riemannian if we assume that the functions $a$ and $b$ are
real, and that the coordinates obey some reality condition. If the
functions and the coordinates are real, the metric will be of
ultra-hyperbolic signature. 

\subsubsection{Case II}

Here we introduce functions $\phi$ and $\psi$ and take
\[
\mathbf{v} = {\partial}_x + {\phi}_y {\partial}_t 
- {\psi}_y {\partial}_y,\qquad
\mathbf{x} = {\partial}_z - {\phi}_t {\partial}_t 
+ {\psi}_t {\partial}_y.
\]
The functions $\phi$ and $\psi$ 
must now satisfy the equations
\bann
{\phi}_{tx} + {\phi}_{yz} 
+ \{ {\phi}_t, {\phi} \} 
+ \{ \psi, {\phi}_y \} = 0,
\\
{\psi}_{tx} + {\psi}_{yz} 
+ \{ {\psi}_t, {\phi} \} 
+ \{ \psi, {\psi}_y \} = 0.
\eann
In terms of these coordinates and functions, the metric is
\[
{\mathbf{g}} = 4 
\left[ \left( dt + \phi_t dz - \phi_y dx \right) \otimes dx +
\left( dy - \psi_t dz +  \psi_y dx \right)  \otimes  dz \right].
\]
In this case, the coordinates $(t, y)$ are again null labelling some of
the null planes in the space. The coordinates $(x, z)$ are not null,
however, and label the \lq\lq rate of change\rq\rq~ of the null planes
(see \cite{NPT} for more on the geometrical interpretation of these
coordinates.)

These equations have been previously investigated 
in connection with {\asd} structures \cite{FP}, 
and are a direct generalisation of the 
Second Heavenly Equation for half-flat 
metrics \cite{Pl}.

\subsubsection{Case III}

Finally, we consider the analogue of the expansion 
used in to reduce the half-flat case to evolution 
form \cite{G1}. We therefore introduce functions 
$g$ and $h$ such that
\[ 
{\mathbf{v}} = {\partial}_t + 
g_y {\partial}_x - h_y {\partial}_z,
\qquad
{\mathbf{x}} = 
- g_t {\partial}_x + h_t {\partial}_z.
\]
We find that the functions $f$ and $g$ 
must satisfy the equations
\be
g_{tt} + \{g_x, g\} + \{h, g_z\} = 0, \qquad
h_{tt} + \{h_x, g\} + \{h, h_z\} = 0.
\ee
In this coordinate system, the metric takes the local form
\[
{\mathbf{g}} = 4 \Delta^{-1} 
\left[ dt \otimes \left( g_t dz + h_t dx \right) +
dy \otimes \left( g_y dz + h_y dx \right) 
- \Delta^{-1} 
\left( g_t dz + h_t dx \right)^2 \right]
\]
with $\Delta = \left( h_t g_y - g_t h_y \right)\,.$
In this system, the coordinates $(t, y)$ still span null planes, however
the geometrical interpretation of the $(x, z)$ coordinates is not 
particularly clear. The main advantage of this coordinate system,
however, is that the equations of motion are in evolution form. This
form of the equations is therefore the natural starting point 
for the study of symmetry algebras \cite{G1}, and the associated 
integrable hierarchy \cite{S1}. We shall therefore concentrate on this
form of the equations from now on.

\section{Integrable Description}

In this section the system (\ref{evol}) will be studied, viewing it as
an integrable system and hence applying various known results from the
theory of integrable systems to it. In particular a Lax pair will be given
for the system, a hierarchy of conservation laws constructed and the
Lie-point symmetry structure calculated. 

\subsection{Lax Pair}

A characteristic feature of an integrable system is the ability to express it as the
compatibility condition for an otherwise overdetermined linear system. To obtain
the system (\ref{evol}) in such a way consider the following vector fields on 
complexified tangent bundle,
$T_c X = TX \otimes \mathbb{C}$ of a four manifold $X\,:$
\[
\begin{array}{rclcrcl}
\mathbf{u} & = & \partial_t\,, && \mathbf{w} & = & \partial_y \,, \\
\mathbf{v} & = & {\partial}_t + 
g_y {\partial}_x - h_y {\partial}_z\,, &&
\mathbf{x} & = &
- g_t {\partial}_x + h_t {\partial}_z\,.
\end{array}
\]
With these define the new vector fields
\begin{eqnarray*}
{\cal L}_0 & = & \mathbf{u} - \lambda \mathbf{x} \,,\\
{\cal L}_1 & = & \mathbf{w} + \lambda \mathbf{v} \,,
\end{eqnarray*}
where $\lambda\in {\mathbb{CP}}^1$ is an auxiliary parameter. The compatibility
conditions for the otherwise overdetermined linear system
${\cal L}_0 \Psi = {\cal L}_1 \Psi = 0 $ results in the following
system of equations:

\bann
&\left[ \mathbf{u}, \mathbf{w} \right] = 0,&
\\
&\left[ \mathbf{u}, \mathbf{v} \right] + 
\left[ \mathbf{w}, \mathbf{x} \right] = 0,&
\\
&\left[ \mathbf{v}, \mathbf{x} \right] = 0.&
\label{compalg}
\eann
With the explicit vector fields given above these reduce to the
system (\ref{evol}):
\begin{equation}
\begin{array}{rcl}
g_{tt} & = & \{g_x, g\} + \{h, g_z\}\,, \\
h_{tt} & = & \{h_x, g\} + \{h, h_z\}\,,
\end{array}
\end{equation}
where, for convenience, the Poisson bracket $\{f,g\}=f_y g_z - f_z g_y$ has
been used. In the previous section the geometry underlying this construction
was given; solutions define a hypercomplex metric on the manifold $X\,.$
Here we just consider the system as an example of a four dimensional integrable
system and study it thus.

\medskip

One interesting reduction of this system is to impose the condition
$g_y=h_z\,.$ One may solve this constraint by introducing a function
$\psi$ such that $h=\psi_y\,,g=\psi_z\,.$ With this it is possible to
integrate (\ref{evol}) and obtain a single evolution equation (\ref{grant})\,.
With this constraint the basic vector fields
$\mathbf{e_i} = (\mathbf{u}\,,\mathbf{v}\,,\mathbf{w}\,,\mathbf{x})\,$
become volume preserving, that is

\[
{\cal L}_{\mathbf{e_i}} {\boldsymbol{\omega}} = 0
\]
where $\boldsymbol{\omega}$ is the volume form
${\boldsymbol{\omega}} = \mathbf{u}\wedge\mathbf{v}\wedge\mathbf{w}\wedge\mathbf{x}\,.$
In this case the metric is hyperK\"ahler rather than hypercomplex, with $\psi$
being related to the K\"ahler potential by a Legendre transformation.
Conversely, one may regard the system (\ref{evol}) as a generalization
of (\ref{grant}) when one relaxes the volume preserving condition on the
vector fields in the associated Lax pair.

\subsection{Formal Solutions}

The system (\ref{grant}) is in Cauchy-Kovaleskaya form, so their formal
solution may be written as a power series in the $t$-variable:

\begin{eqnarray*}
h & = & \sum_{n=0}^\infty h_n(x,y,z) t^n \,,\\
g & = & \sum_{n=0}^\infty g_n(x,y,z) t^n
\end{eqnarray*}
and the differential equations reduce to recursion relations between the
coefficients $h_n\,,g_n\,.$ Thus a formal solution may be derived from the
coefficients $(g_0\,,h_0\,,g_1\,,h_1)\,,$ or equivalently, in terms of the initial
data $\left.(g\,,h,g_t\,,h_t)\right|_{t=0}$ on the $t=0$ hypersurface in $X\,.$
This shows that the general solution depends on four arbitrary functions of three
variables.

\bigskip

One simple, but explicit, solution may be obtained from the ansatz

\begin{eqnarray*}
g & = & t y + G(t,x,z) \,,\\
h & = & t\,.
\end{eqnarray*}
With this the nonlinearites in (\ref{grant}) disappear and one is left with
a linear equation for $G\,,$ which after a simple change of variable, is just
the three dimensional Laplace equation. Other simple solutions may be obtained
by taking known hypercomplex metrics and reexpressing them in the above form.
Some examples of solutions obtained in this way will be given later.

\subsection{Symmetry Structure}

While symmetry techniques may be applied to any system of differential
equations, the Lie-point symmetries of integrable systems have a particularly
rich structure compared to non-integrable systems. Indeed, possible integrable
systems may often be indentified by an increase in the dimension of the
Lie-algebra of symmetries, as compared to nearby non-integrable systems.

\bigskip

Let $x=(x_1\,,\ldots\,,x_p)$ and $u=(u^1\,,\ldots\,,u^q)$ be sets of independent
and dependent variables, and consider a set of differential equations of
degree $k$, given by
\[
\Delta^i(x,u^{(k)}) = 0 \,, \qquad\qquad i=1\,,\ldots\,,m\,.
\]
Lie-point symmetries are generated by the vector field
\[
{\mathbf{v}} = \sum_{i=1}^p \xi_i(x,u) \frac{\partial~}{\partial x_i}+
\sum_{\alpha=1}^q \phi_\alpha(x,u) \frac{\partial~}{\partial u^\alpha}\,,
\]
where the coefficients are determined by the criterion
\[
\left.pr^{(k)} {\mathbf{v}} (\Delta)\right|_{\Delta=0} = 0
\]
where $pr^{(k)}{\mathbf{v}}$ is the $k$-th prolongation of the vector field
${\mathbf{v}}\,.$ These ideas, and notation, are standard, see \cite{O}.

\bigskip

To apply such a procedure to the system ({\ref{evol}) it is convenient to convert the
system from a second order system in two independent variables to a first order
system in four independent variables by
introducing potentials and integrability conditions. Therefore let:
\[
A = g_t, B= g_y, C = h_t, D = h_y, 
\]
in which (\ref{evol}) becomes
\begin{equation}
\begin{array}{rcl}
A_y - B_t & = & 0,
\\
C_y - D_t & = & 0,
\\
A_t + A_x B - B_x A + C B_z - D A_z & = & 0,
\\
C_t + C_x B - D_x A + C D_z - C_z D & = & 0.
\end{array}
\label{frst}
\end{equation}
Since these are first order the calculation of the first prolongation of the
vector field is easy, the evolutionary form of the system also giving a
distinguished variable $t$ to eliminate in the course of the calculations.


The result is that we get five families of symmetries, the
vector
fields which generate these families being (where $k$ is a constant,
$\phi, \psi$ are functions of the coordinate
$y$, and $a_1, a_2$ are functions of coordinates $(x, z)$):
\[
\begin{array}{rcl}
{\mathbf{v}}^1 [k] &=& 
k 
\left( 
- t \partial_t + 2A \partial_A + B \partial_B 
+ 2C \partial_C + D \partial_D 
\right),
\\
{\mathbf{v}}^2 [{\phi}] &=& 
- \phi \partial_t 
+ A \phi_y \partial_B 
+ C \phi_y \partial_D,
\\
{\mathbf{v}}^3 [{\psi}] &=& 
- t \psi_y \partial_t 
- \psi \partial_y
+ 
\psi_y 
\left( 
A \partial_A + B \partial_B 
+ C \partial_C + D \partial_D 
\right) 
+ 
t \psi_{yy} 
\left( A \partial_B + C \partial_D \right),
\\
{\mathbf{v}}^4 [{a_1}] &=& - a_1 \partial_x 
- \left( a_{1x} A - a_{1z} C \right)
\partial_A - \left( a_{1x} B - a_{1z} D \right) \partial_B,
\\
{\mathbf{v}}^5 [{a_2}] &=& - a_2 \partial_z 
- \left( a_{2z} C - a_{2x} A \right)
\partial_C - \left( a_{2z} D - a_{2x} B \right) \partial_D.
\end{array}
\]

\noindent The first of these generators simply generates a scaling symmetry of the
equations. Therefore, if $A, B, C, D$ constitute a solution of the
equations, then so do ${\tilde A}, {\tilde B}, {\tilde C}, {\tilde D}$
defined by 
\[
\begin{array}{rcl}
{\tilde A} (t, x, y, z) &=& {\rm e}^{2k} A ( {\rm e}^{k} t, x, y, z),
\\
{\tilde B} (t, x, y, z) &=& {\rm e}^{k} B ( {\rm e}^{k} t, x, y, z),
\\
{\tilde C} (t, x, y, z) &=& {\rm e}^{2k} C ( {\rm e}^{k} t, x, y, z),
\\
{\tilde D} (t, x, y, z) &=& {\rm e}^{k} D ( {\rm e}^{k} t, x, y, z).
\end{array}
\]
Similarly, the second generator generates a translation in the $t$
coordinate along with a redefinition of fields, so that in this case
\[
\begin{array}{rcl}
{\tilde A} (t, x, y, z) &=& A ( t + \phi, x, y, z),
\\
{\tilde B} (t, x, y, z) &=& 
B ( t + \phi, x, y, z) + \phi_y A ( t + \phi, x, y, z),
\\
{\tilde C} (t, x, y, z) &=& C ( t + \phi, x, y, z),
\\
{\tilde D} (t, x, y, z) &=& 
D ( t + \phi, x, y, z) + \phi_y C ( t + \phi, x, y, z).
\end{array}
\]
These are the only symmetries that it is possible to generally
exponentiate explicitly. The non-zero commutators for this
Lie algebra are:
\[
\begin{array}{rcl}
\left[ {\mathbf{v}}^1 [k], {\mathbf{v}}^2 [\phi] \right] 
&=& {\mathbf{v}}^2 [k\phi],
\\
\left[ {\mathbf{v}}^2 [\phi], {\mathbf{v}}^3 [\psi] \right] 
&=& {\mathbf{v}}^2 [\psi\phi_y - \phi\psi_y],
\\
\left[ {\mathbf{v}}^3 [\psi], {\mathbf{v}}^3 [\chi] \right] 
&=& {\mathbf{v}}^3 [\chi\psi_y - \psi\chi_y],
\\
\left[ {\mathbf{v}}^4 [a], {\mathbf{v}}^4 [b] \right] 
&=& {\mathbf{v}}^4 [a_x b - a b_x],
\\
\left[ {\mathbf{v}}^4 [a], {\mathbf{v}}^5 [b] \right] 
&=& {\mathbf{v}}^4 [a_z b] - {\mathbf{v}}^5 [a b_x],
\\
\left[ {\mathbf{v}}^5 [a], {\mathbf{v}}^5 [b] \right] 
&=& {\mathbf{v}}^5 [a_z b - a b_z],
\end{array}
\]
where $k, l$ are arbitrary constants, $\psi, \phi$ are arbitrary
functions of $y$, and $a, b$ are arbitrary functions of $x$ and $z$.

\bigskip

From the structure of these commutators one may decompose the Lie
algebra $L$ generated by these vector fields into a direct sum
$L=L_1 \oplus L_2\,,$ where
\begin{eqnarray*}
L_1 & = & \{ {\mathbf{v}}^1[k],{\mathbf{v}}^2[\phi],{\mathbf{v}}^3[\psi]\}\,,\\
L_2 & = & \{ {\mathbf{v}}^4[a],{\mathbf{v}}^5[b]\}\,.
\end{eqnarray*}
The sub-Lie-algebra $L_1$ decomposes further as a semidirect product
$L_1 = S \triangleright R$ where
\begin{eqnarray*}
S & = & \{ {\mathbf{v}}^3[\psi]\}\,,\\
R & = & \{ {\mathbf{v}}^1[k],{\mathbf{v}}^2[\phi]\}\,.
\end{eqnarray*}

\noindent The subalgebra $L_2$ is isomorphic to vector fields on a $2$-dimensional
surface and correspond to coordinate transformation in the $x\,,z$-variables.
This is to be expected, since in the hyper-K\"ahler case the
vectors are all divergence free and the symmetries turn out to be
related to symplectic diffeomorphisms of two dimensional planes. In our
case, we have simply dropped the divergence-free condition from the
vector fields, and the symmetry group becomes related to the larger
group of diffeomorphisms, since there is no natural symplectic structure
any more. Similarly the vector fields ${\mathbf{v}}^2$ generate coordinate
transformations and the vector fields ${\mathbf{v}}^1$ generate constant
rescalings of the metric. The only vector fields which generate genuinely
new metrics are those in $S\,.$

\subsection{Conservation Laws}

In this section a hierarchy of conservation laws of the form
\[
g^{{\mu\nu}} \nabla_\mu j_{\nu}^{(n)} = 0 \,,\qquad n=0\,,1\,, \ldots
\]
will be constructed. This expression is clearly covariant, but in the
calculations it will be necessary to use a particular form of the
metric and the associated field equations. Explicitly we consider
metrics of the form
\begin{equation}
{\mathbf{g}} =2 \Delta^{-\frac{1}{2}}
[dt \otimes (a_t dz+b_t dx) + dy \otimes (a_y dz + b_y dx)]
\label{detoneheaven}
\end{equation}
where $\Delta=(a_t b_y - a_y b_t)$ and with $a$ and $b$ being solutions of the field equations
(\ref{heaven})\,. The conformal fact in (\ref{detoneheaven})
has been fixed so that $\det g_{ij}=1\,;$ such a fixing does not
change the hypercomplex or Hermitian properties of the metric.
One obvious extension of these results would be to introduce the
notion of a conformally invariant conservation law.

\bigskip

The starting point of this construction, a generalization of a procedure
first applied to nonlinear $\sigma$-models \cite{BIZZ,LP}, is the solution
$\Psi$ to the Lax pair
\begin{equation}
\begin{array}{rcl}
\left[\lambda \partial_t + a_t\partial_x - b_t \partial_z\right]\Psi & = & 0\,, \\
\left[\lambda \partial_y + a_y\partial_x - b_x \partial_z\right]\Psi & = & 0\,,
\end{array}
\qquad \lambda\in{\mathbb{CP}}^1\,.
\label{conlawlax}
\end{equation}
Expanding $\Psi$ as a power series in $\lambda\,,$ i.e. as
$\Psi=\sum_{n=0}^\infty \lambda^n \Psi_n$ and equating coefficients
yields the following equations for the $\Psi_0$-term:
\begin{equation}
\begin{array}{rcl}
\left[a_t \partial_x - b_t \partial_z \right]\Psi_0 & = & 0 \,,\\
\left[a_y \partial_x - b_y \partial_z \right]\Psi_0 & = & 0 \,,
\end{array}
\label{lowest}
\end{equation}
and the recursion relations
\begin{equation}
\begin{array}{rcl}
\partial_t \Psi_n & = & (-a_t\partial_x + b_t \partial_z ) \Psi_{n+1}\,, \\
\partial_y \Psi_n & = & (-a_y\partial_x + b_y \partial_z ) \Psi_{n+1}\,.
\end{array}
\label{recursion}
\end{equation}
The first set of equations imply that $\Psi_0=\Psi_0(t,y)$ and so we take the
seed solution to be
\[
\Psi_0 = \left( \begin{array}{rcl} t \\ y \end{array} \right)
\]
(here we have assembled two independent solutions into a vector). This seed
solution will generate, via the recursion relations (\ref{recursion}) the
full solution to the Lax pair (\ref{conlawlax}). This function defines the
twistor surfaces in the corresponding twistor space. Another family of conservation
laws may be obtained starting from the expansion 
${\tilde\Psi}=\sum_{n=0}^\infty \lambda^{-n} {\tilde\Psi}_n$ and the relationship
between $\Psi$ and $\tilde\Psi$ on the equator of ${\mathbb{CP}}^1$ defines the
twistor space, via a patching construction \cite{NPT,S2}.

\bigskip

\begin{proposition}

The currents $j_{\mu}^{(n)}$ defined by

\[
\begin{array}{rclcrcl}
j_t^{(n)} & = & 0 \,, & \quad & j_x^{(n)} = \Delta^{\frac{1}{2}} \, \partial_x \Psi_{n+1}\,,
\\
j_y^{(n)} & = & 0 \,, & \quad & j_z^{(n)} = \Delta^{\frac{1}{2}} \, \partial_z \Psi_{n+1}\,,
\end{array}
\]
are conserved.

\end{proposition}

\bigskip

\noindent{\bf Proof} With the particular metric (\ref{detoneheaven})

\begin{eqnarray*}
g^{\mu\nu} \nabla_\mu j_{\nu}^{(n)} & = &
\partial_t[ - a_y \partial_z \Psi_{n+1} + b_y \partial_z \Psi_{n+1}]
+\partial_y[ + a_t \partial_z \Psi_{n+1} - b_t \partial_z \Psi_{n+1}]\,,\\
& = & \partial_t [ \partial \Psi_n ] - \partial_y [ \partial_t \Psi_n ] \,,\\
& = & 0 \,.
\end{eqnarray*}
This proof uses the condition $\det g_{ij}=1\,,$ so the Christoffel symbols
$\Gamma_{\nu\mu}^\mu=0\,.$

\medskip

\endproof

\section{Comments}

Underlying the integrability of the multidimensional systems presented here is
the existence of a twistor space. This paper has, though, only concentrated on the field
equations and the associated Lax pairs with little mention of the properties of the
corresponding twistor space -- in terms of a double fibration

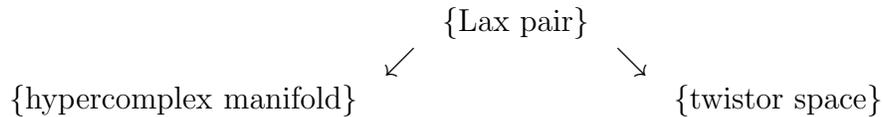
\[
\begin{array}{rclcl}
& &
\left\{ {\rm Lax~pair} \right\} & & \\
& \swarrow & & \searrow & \\
\left\{ {\rm hypercomplex~manifold}\right\} & & & &
\left\{ {\rm twistor~space} \right\}
\end{array}
\]
we have said little about about the structure of the right hand side. Such twistor space
have the special property that they fibre over ${\mathbb{CP}}^1\,$ \cite{Bo}, unlike those
for more general \asd\ Weyl spaces or scalar flat K\"ahler spaces. This is manifested in the
simple $\lambda$-dependence in the Lax pairs for hypercomplex manifolds -- the Lax pairs
for scalar flat K\"ahler \cite{Pa} and general \asd\ Weyl spaces \cite{G2,MW} involve
$\partial_\lambda$-terms. The hypercomplex manifolds studied here come from the
conformally invariant condition $d{\mathbf{A}}=0\,,$ and so one would expect the
corresponding twistor space to exhibit certain extra properties. General
hypercomplex manifolds (without this condition) will be studied in the sequel to this
paper, this also containing the connection between the approach developed here and the
Obata connection.

\bigskip

One characteristic feature of integrable systems is the existence of an associated
hierarchy. Such hierarchies may be constructed by studying the generalized symmetry
structure of the original systems of equations \cite{O}. Such hierarchies have been
constructed for hyperK\"ahler metrics in \cite{S2}. It remains to see how such ideas
may be extended to the hypercomplex systems studied here.

\section*{Acknowledgments}

Financial support was provided by the EPSRC.

\section*{Appendix}

While many of the ideas in this paper will generalize to $4N$-dimensional manifolds, when
$N=1$ another generalization is possible. Underlying the integrability of the structures
studied in this paper is the existence of a suitable twistor space, and hypercomplex
manifolds automatically have such twistor spaces. However in four dimensions the existence
of a twistor space follows from the Weyl tensor being anti-self-dual, and the hypercomplex
condition here implies, not is implied by, this condition. Thus a possible generalization
is to study metrics with anti-self-dual Weyl tensor and which are also scalar flat. Such
a system has an other reduction to scalar flat K\"ahler metrics. These different systems
and their interconnections are summarized in the following diagram:

\bigskip

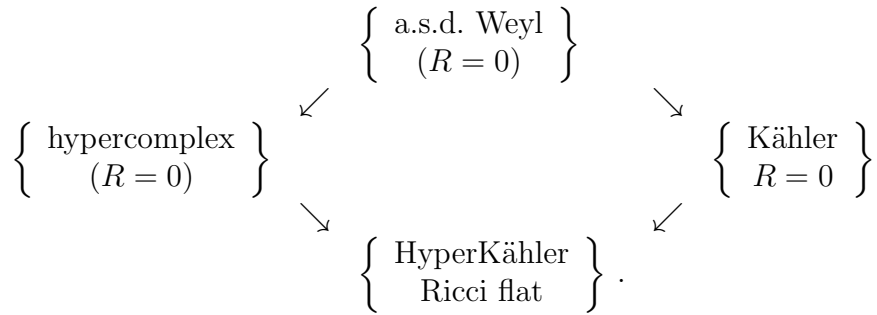
\[
\begin{array}{rclcl}
& &
\left\{ \begin{array}{c} {\rm a.s.d.~Weyl} \\ (R=0) \end{array}\right\} & & \\
& \swarrow & & \searrow & \\
\left\{ \begin{array}{c} {\rm hypercomplex} \\ (R=0) \end{array}\right\} & & & &
\left\{ \begin{array}{c} {\rm K\ddot{a}hler} \\ R=0 \end{array} \right\}\\
& \searrow & & \swarrow & \\
& &
\left\{ \begin{array}{c} {\rm HyperK\ddot{a}hler} \\ {\rm Ricci~flat}
\end{array}\right\}\,. & &
\end{array}
\]
The conditions $(R=0)$ in brackets indicate how the conformal factor for otherwise
conformally invariant conditions have been fixed.

\bigskip

An analogous system of equations to (\ref{grant}) for the scalar-flat, anti-self-dual-Weyl
systems is given by

\[
\begin{array}{rcl}
g_{tt}+(e^{-\psi} \{g,h\})_z & = & e^{-\psi} \{g,h_z-g_y\} \,, \\
h_{tt}+(e^{-\psi} \{g,h\})_y & = & e^{-\psi} \{h,h_z-g_y\} \,, \\
(e^{+\psi})_{tt} +\{g,\psi_y\} - \{h,\psi_z\} & = & 0 
\end{array}
\]
(these certainly imply the geometric conditions though whether they are implied by
them is unclear). The corresponding metric is given by
\[
{\mathbf{g}} =
+2 dy \Bigg\{ h_t dt + h_x dx - \frac{h_t e^\psi}{\Delta} [ h_t dy + g_t dz]\Bigg\} 
+2 dz \Bigg\{ g_t dt + g_x dx - \frac{g_t e^\psi}{\Delta} [ h_t dy + g_t dz]\Bigg\}\,.
\]
where $\Delta=h_t g_x - g_t h_x\,.$
The two reductions above are easy to see from this system:

\begin{itemize}
\item 
when $\psi=0\,,$ this system reduces to 

\begin{eqnarray*}
g_{tt} & = & \{h,g_z\}-\{g,g_y\} \,, \\
h_{tt} & = & \{h,h_z\}-\{g,h_y\} \,,
\end{eqnarray*}
that is, to the hypercomplex systems studied in the main body of this paper.
The further reduction $h=\theta_y\,,g=\theta_z$ reduces this down to
the hyperK\"ahler equation (\ref{grant})\,.

\item imposing the K\"ahler condition on this system
gives $h=\theta_y\,,g=\theta_z$ and the
first set of equations simplify to

\begin{eqnarray*}
\theta_{tt} + e^{-\psi} \{\theta_z,\theta_y\} & = & 0 \,, \\
(e^{+\psi})_{tt} +\{\theta_z,\psi_y\}-\{\theta_y,\psi_z\} & = & 0 \,.
\end{eqnarray*}
These are the analogues of the well-known scalar flat K\"ahler equations
\cite{F}, written in evolutionary form.
It is easy to
find solutions, such as the one which gives the Burns metric. The further
reduction $\psi=0$ reduces this down to
the hyperK\"ahler equation (\ref{grant})\,.
\end{itemize}

\noindent One interesting class of solutions to all these systems comes
from imposing an $SU(2)$ symmetry on the metrics. Such metrics are often
referred to as Bianch IX metrics. This symmetry reduces the field equations
from partial differential equations down to systems of coupled ordinary
differential equations which may be integrated directly. These ideas
may also be applied to other Bianchi metrics.

\end{document}